\documentclass[twocolumn]{aastex63}

\usepackage{graphicx}

\begin{document}

\title{Carbon-grain sublimation: a new top-down component of protostellar chemistry}

\correspondingauthor{Merel L.R. van 't Hoff}
\email{mervth@umich.edu}

\author{Merel L.R. van 't Hoff}
\affil{Department of Astronomy, University of Michigan, 500 Church Street, Ann Arbor, MI 48109, USA}

\author{Edwin A. Bergin}
\affil{Department of Astronomy, University of Michigan, 500 Church Street, Ann Arbor, MI 48109, USA}

\author{Jes K. J{\o}rgensen}
\affil{Niels Bohr Institute, University of Copenhagen, {\O}ster Voldgade 5-7, 1350 Copenhagen K., Denmark}

\author{Geoffrey A. Blake}
\affil{Division of Geological \& Planetary Sciences, MC 150-21 California Institute of Technology Pasadena, CA 91125, USA}

% ======================================================================
% ABSTRACT 
% ======================================================================

\begin{abstract} 

\noindent Earth's carbon deficit has been an outstanding problem in our understanding of the formation of our Solar System. A possible solution would be the sublimation of carbon grains at the so-called soot line ($\sim$300 K) early in the planet-formation process. Here, we argue that the most likely signatures of this process are an excess of hydrocarbons and nitriles inside the soot line, and a higher excitation temperature for these molecules compared to oxygen-bearing complex organics that desorb around the water snowline ($\sim$100 K). Such characteristics have been reported in the literature, for example, in Orion KL, although not uniformly, potentially due to differences in observational settings and analysis methods of different studies or related to the episodic nature of protostellar accretion. If this process is active, this would mean that there is a heretofore unknown component to the carbon chemistry during the protostellar phase that is acting from the top down -- starting from the destruction of larger species -- instead of from the bottom up from atoms. In the presence of such a top-down component, the origin of organic molecules needs to be re-explored. 

\vspace{1cm}

\end{abstract}

\keywords{}

% ======================================================================
% INTRODUCTION
% ======================================================================

\section{Introduction} \label{sec:intro}

One of the main goals in the fields of exoplanets and planet formation is to determine the composition of terrestrial, potentially habitable, planets and to link this to the composition of protoplanetary disks. A longstanding puzzle in this regard is the Earth's severe carbon deficit. Earth is four orders of magnitude depleted in carbon compared to interstellar grains, and two well-characterized comets 1P/Halley and 67P/Churyumov-Gerasimenko (hereafter, 67P) (e.g., \citealt{Geiss1987,Bergin2015,Rubin2019}). The exact depletion is uncertain as a significant amount of carbon could be present in the Earth's core, but even genereous upper limits suggest one to two orders of magnitude depletion for the bulk Earth (see e.g., \citealt{Marty2012,Fischer2020}; Li et al. 2020, subm.). A similar amount of depletion is seen in CI chondrites that are thought to represent the most primitive material in the Solar System and otherwise reflect solar abundances in terms of composition \citep{Wasson1988}. Moreover, this problem exists beyond the Solar System as carbon deficits in polluted white dwarfs are indicative of the accretion of carbon-depleted rocky material \citep[e.g.,][]{Jura2006}.

The only solution to this conundrum is that in the inner few au of planet-forming systems, carbon has to be in the gas phase instead of the refractory phase, and as such, becomes unavailable for accretion onto rocky bodies. Thus, there must be a mechanism to destroy carbon grains while leaving silicate grains intact. Furthermore, this must happen prior to planetesimal formation as it is much easier to destroy smaller grains than it is to break apart planetesimals and entirely ablate them. This points towards the early, embedded, phases in the evolution of young stars before significant grain growth sets in (i.e., the Class 0 and early Class I protostellar stages). As this mechanism is central to the supply of carbon to terrestrial worlds, constraining it is of fundamental importance. Carbon-grain destruction has been explored in the literature with a primary focus on oxidation \citep[e.g.,][]{Finocchi1997,Lee2010,Gail2017}, but detailed models by \citet{Anderson2017} and \citet{Klarmann2018} suggest that this mechanism is ineffective.

Here, we focus on sublimation of refractory carbon grains at a location that can be labeled as the ``soot line'' \citep{Kress2010}. Although the specific molecular form of carbon in grains is unknown, the majority of refractory carbon-rich solids are sublimating at temperatures between $\sim$350 and 450 K (\citealt{Nakano2003}; \citealt{Gail2017}; Li et al. 2020, subm.). If the process of carbon-grain sublimation is indeed active, this would mean that there is a heretofore unrecognized contributor to the rich carbon chemistry. This pathway acts from the top down, that is, starting from the destruction of larger molecules, instead of from the bottom up as in traditional gas and ice chemistry (see also \citealt{Tielens2011} for a discussion on top-down chemistry). In Sect.~\ref{sec:Signatures}, we outline what the observational signatures of carbon-grain sublimation could be. In Sect.~\ref{sec:Evidence} we review whether there is current evidence that this process is happening and in Sect.~\ref{sec:Discussion} we discuss future steps to establish whether carbon-grain sublimation is a common process during star- and planet formation.

%............. Figure: Cartoon ..............................................
\begin{figure*}[ht!]
\centering
\includegraphics[width=\textwidth]{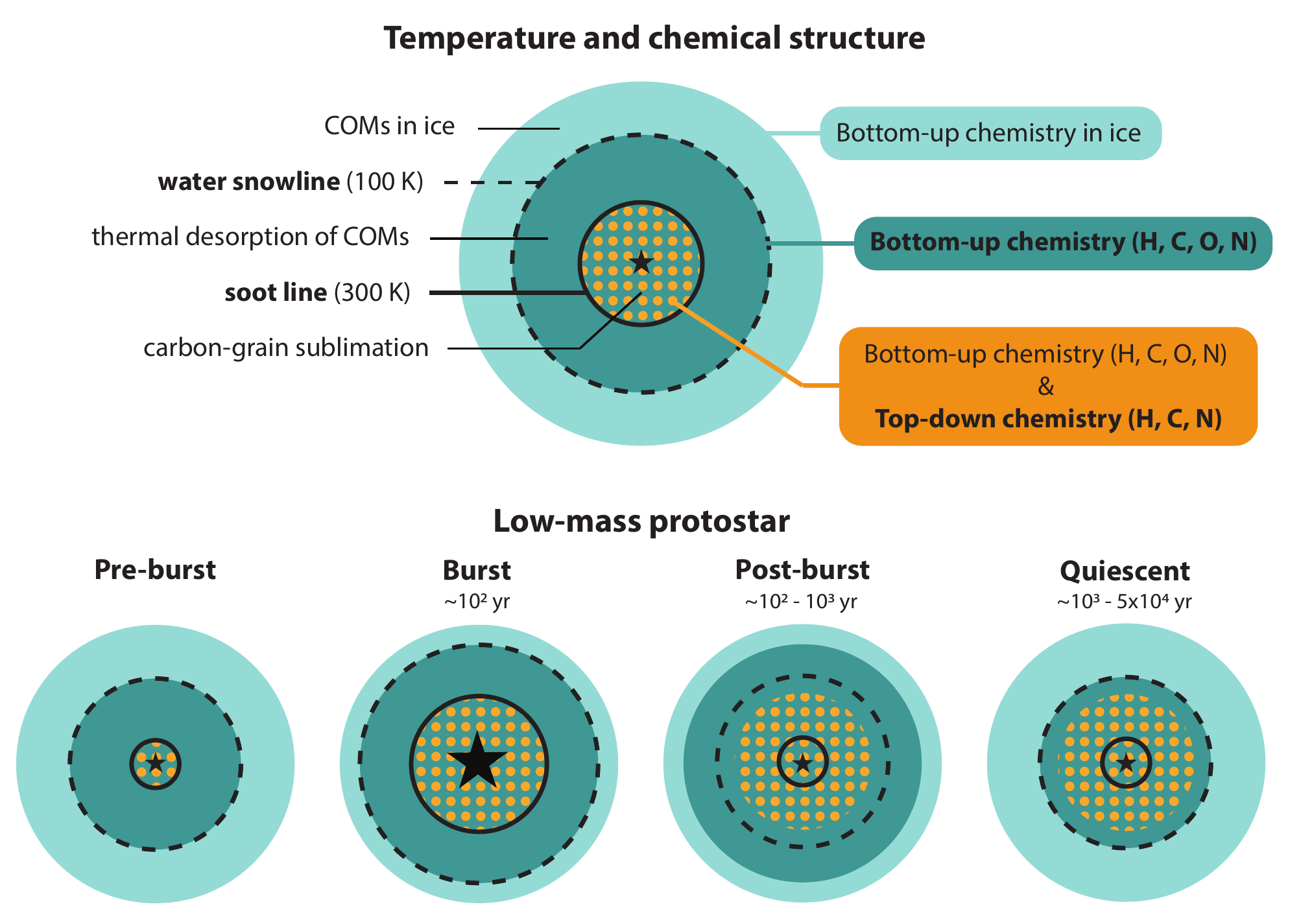}
\caption{\textit{Top panel:} Schematic of the temperature and chemical structure around a protostar (not to scale). In the outer most part of the envelope, complex organic molecules (COMs) are present in the ice, where they have been formed, for example, through a sequence of hydrogenation reactions starting with the hydrogenation of CO (that is, bottom up). When the temperature exceeds $\sim$100 K at the water snowline (dashed black line; a few 10s of au for 1 $L_\odot$), the COMs desorb off the dust grains and may continue to react in the gas phase. Inside the soot line ($\sim$300 K at a few au for 1 $L_\odot$; solid black line), carbon grains sublimate, providing a top-down component to the chemistry (depicted with orange dots) that enriches the gas with carbon and nitrogen. \textit{Bottom panels:} Schematic of the evolution of the temperature and chemistry for a low-mass protostar that undergoes an accretion burst, using the same lines and color coding as in the \textit{top panel}. Depending on the strength of the burst, the soot line may or may not shift past the quiescent snowline location. In the scenario illustrated here, the soot line does not shift past the quiescent snowline. The duration of the different phases is indicated above the panels. See Sect.~\ref{sec:Tex} for details.} 
\label{fig:Cartoon}
\end{figure*}
%....................................................................................

% ======================================================================
% 2. SIGNATURES 
% ======================================================================

\section{Signatures of carbon-grain sublimation and top-down chemistry} \label{sec:Signatures}

In hot cores, it is common to think of molecular abundance changes across the water snowline ($\sim$100 K) as many complex molecules have binding energies similar to water. A simple approach would therefore be to look for an additional change in the chemical structure at the temperature of carbon-grain sublimation (see Fig.~\ref{fig:Cartoon}, top panel). The questions then become what these signatures are and where to look for them. 

Before we explore these issues we need to address the sublimation temperature of carbonaceous grains, which is $\sim$425 K for pressures representing the regions in the solar nebula disk at a few au from the Sun (i.e., the formation zone for meteoritic material) (\citealt{Nakano2003}; \citealt{Gail2017}; Li et al. 2020, subm.). This number is based upon the recent analysis of Li et al. (2020, submitted) who note that meteoritic constraints bound the sublimation temperature of the main carbon carrier to be within 200 - 650~K.  Drawing upon the laboratory experiments of cometary organic analogs of \citet{Nakano2003} we adopt their number of 425~K. Sublimation has a well characterized exponential relation between pressure and temperature (e.g., the Clapeyron-Classius equation). To obtain a rough estimate of the sublimation temperature at hot core pressures we use the dP$_{vap}$/dT (where P$_{vap}$ is the vapor pressure) relations of PAHs characterized in the lab \citep{Goldfarb08, AslamSiddiqi10} and assume that only the density of H$_2$ and temperature changes (i.e., abundance is constant).  Based on these dependencies \citep[see also,][]{Bergin2018}, we obtain a (very) rough estimate of $\sim$300~K for the temperature of carbon-grain destruction at reduced pressures (densities of $10^7-10^8$ H$_2$ cm$^{-3}$, 300~K; P $\sim 10^{-13}$~bar) in the inner region around protostars, compared to the inner disk \citep[P(1~au) $\sim$ 10$^{-6}$ bar;][]{DAlessio05}. We note that strictly speaking the above relations refer to pure ices and the temperature of the medium. In this case the temperature refers to the gas and dust temperature which are coupled in these regions.

\subsection{Excess carbon and nitrogen}

The most obvious effect of carbon-grain sublimation is a flooding of the gas with carbon. Since most of the oxygen is expected to be locked up in water \citep{Bergin2002,Caselli2012}, this carbon enters gas that is rich in hydrogen, nitrogen (mostly in N$_2$), CO, and H$_2$O. Carbon will therefore most likely react to form hydrocarbons and nitriles (molecules with a C$\equiv$N bond, such as HCN and CH$_3$CN) as shown in chemical models of protoplanetary disks in which the destruction of carbon grains is simulated by an excess of C$^+$ \citep{Wei2019}. This gets complicated by the fact that some hydrocarbons and nitriles also form bottom up in the gas or ice. 

However, based on the cometary inventory, there is significantly more nitrogen contained in refractories than in volatile ices \citep{Rice2018,Rubin2019}. Carbon is similar to nitrogen in comet 67P, and both are distinct from oxygen, which is twice as abundant in volatile ices as in refractory organics \citep{Rubin2019}. The refractory C/N elemental ratio in carbonaceous chondrites, comets 1P/Halley and 67P is $\sim$5-30 \citep{Bergin2015,Jessberger1988,Rubin2019}. Assuming 50\% of the elemental carbon is in refractory form (150 ppm with respect to H; \citealt{Mishra2015}) then the destruction of carbon grains releases 5-30 ppm of refractory nitrogen into the gas. This is about 10-50\% of the present day cosmic nitrogen abundance \citep{Nieva2012}. 

Recently, evidence of the presence of ammonium salts was reported in comet 67P, providing an additional nitrogen-bearing component \citep{Altwegg2020,Poch2020}. These works derive an upper limit of 40 wt\% for the mass fraction of ammonium salts, but depending on the composition of the salts, a 10 wt\% mass fraction can be enough for a solar C/N ratio ($\sim$3.4, \citealt{Lodders2010}) for the entire comet. In these cases, $\sim$10\% to almost 40\% of all nitrogen, respectively, would be in refractories. Sublimation of carbon grains with these compositions would then release 9-36 ppm of refractory nitrogen. Furthermore, the desorption temperature of ammonium salts (roughly 200-250 K, \citealt{Clementi1967,Raunier2004,Bossa2008,Danger2011,Vinogradoff2011,Noble2013,Bergner2016}) is higher than that of water and comparable to that of refractory carbon. In the outlined scenarios, $\sim$60--90\% of all nitrogen is expected to be in ammonium salts, so these salts could deliver an additional amount of nitrogen to the gas inside the soot line. 

Sublimation of carbon grains would thus lead to an excess of carbon and nitrogen in the gas phase (see Fig.~\ref{fig:Cartoon}, top panel). The exact form in which these elements will be released (e.g., N along with C or NH with CH or C with CN or larger fragments) is unknown. If polycyclic aromatic hydrocarbons (PAHs) are freed, this carbon will not be readily available for gas-phase chemistry. However, both low and high-mass protostars lack PAH emission \citep{vanDishoeck2000,Geers2009}, while according to radiative transfer models, PAH emission should always be detectable in Herbig Ae/Be disk/envelope systems if they are present \citep{Manske1999}. It seems therefore unlikely that the 90\% of carbon that is missing in inner Solar System bodies is present in PAHs. Another unknown is which molecules will be formed exactly and in what amounts. However, cometary compositions indicate that the top-down signal could dominate over the bottom-up contribution. Finally, it is possible that a highly refractory component might exist (e.g. SiC), but the carbon deficit in even the most primitive meteorites requires significant carbon grain destruction to the order of 90\% \citep{Bergin2015} at the time these materials are isolated from the gaseous nebula.

\subsection{Isotopic fractionation}

In the ISM, only $< 17 \pm 11$\% of all nitrogen is contained in refractories, while this is $20-85$\% in comets \citep{Jensen2007,Bergin2015,Rice2018,Rubin2019}. At some point during the star-formation process, nitrogen must thus be captured and placed into refractory material. The detection of benzonitrile in dark clouds suggest that there may be a pathway to do this at low temperatures \citep{McGuire2018}. Signatures of low-temperature formation, such as isotope fractionation, may therefore be present in the nitrogen that comes free upon grain sublimation. This may distinguish molecules formed through this top-down process from species formed in the gas through a bottom-up pathway. However, complex nitrogen-bearing molecules formed in the ice may carry similar fractionation features. 

\subsection{Higher excitation temperature for N-COMs}\label{sec:Tex}

Observing the predicted excess of hydrocarbons and nitriles inside the soot line would require spatially resolving the $\gtrsim 300$ K region. Protostars are the best sources to target because 1) the sublimation has to happen early in the planet-formation process and 2) the soot line is at larger distances than in protoplanetary disks due to the higher accretion rates of these younger systems and their different density structures that result in less shielding of the stellar irradiation. If the soot line is not resolved, a comparison of the excitation temperatures of nitrogen-bearing and oxygen-bearing complex organic molecules (N-COMs and O-COMs, respectively) could still indicate whether carbon-grain sublimation is taking place: the excitation temperatures of N-COMs, enhanced at temperatures $\gtrsim 300$ K, will be higher than those of O-COMs that will be uniformly present inside the water snowline ($\gtrsim 100$ K; Fig.~\ref{fig:Cartoon}, top panel). In case of a large abundance of N-COMs formed through bottom-up chemistry outside the snowline, N-COM emission may be characterized by a hot ($\gtrsim 300$ K) and cold component. However, whether a temperature signal can be observed will depend on the extent of the $\gtrsim 300$ K region within the telescope beam, and is thus not as good a diagnostic as spatial differentiation. 

A separate complication is that the temperature profile itself will change as the young star evolves. Accretion is episodic (see e.g., \citealt{Hartmann1996,Evans2009,Scholz2013}) and enhanced accretion rates cause the luminosity and thus the temperature to increase. This leads to chemical changes due to the soot line and snowlines moving outward (e.g., \citealt{Lee2007,Visser2012} and see Fig~\ref{fig:Cartoon}, burst panel). Catching a protostar during such a burst phase would thus make the detection of grain-sublimation signatures easier. Post-burst, the dust and gas temperature rapidly decay ($\sim$1 year) to levels associated with the quiescent protostellar luminosity \citep{Johnstone2013}. However, chemical changes persist and sublimated molecules remain in the gas phase in the inner envelope for a depletion timescale of $\sim$10$^2-10^3$ years \citep{Lee2007,Visser2012}. Thus it is possible that a chemical signature of carbon-grain sublimation is present even in cases where the temperature signature is absent (see Fig.~\ref{fig:Cartoon}, post-burst panel). 

If the soot line does not get shifted past the quiescent snowline location, the species formed inside the extended soot line will remain present in the gas phase (see Fig.~\ref{fig:Cartoon}, quiescent panel). Alternatively, if the burst pushes the soot line beyond the quiescent snowline location, newly formed species outside the quiescent snowline will also freeze out. In this scenario there will be no spatial differentiation between gas-phase N-COMs and O-COMs as they both extend out to the quiescent snowline location. During the quiescent phase, it will thus depend on the burst location of the soot line compared to the quiescent lcoation of the snowline whether there is a spatial difference between N-COMs and O-COMs. If the spatial difference is small or absent, the only signal of carbon-grain sublimation may be an enhanced abundance of N-COMs. Whether a clear signature can be observed for an individual protostar will thus depend on the time since its last accretion burst and the strength of that, and potentially previous, bursts.

% ======================================================================
% 3. OBSERVATIONAL EVIDENCE
% ======================================================================

\section{Observational evidence for carbon-grain sublimation and top-down chemistry} \label{sec:Evidence}

An increased abundance of hydrocarbons and nitriles inside the soot line, a smaller spatial extent for these species compared to oxygen-bearing COMs and/or higher excitation temperatures than O-COMs are thus expected signatures for sublimation of carbon grains. A clear manifestation of these signatures may be complicated by the occurrence of accretion bursts and the presence of N-COMs formed through bottom-up chemistry. Is there any evidence in the literature that this process is indeed taking place?

\subsection{N-COM/O-COM spatial distribution and abundance correlation}\label{sec:ObsSpatDis}

Probably the most well-known example of N-COMs tracing different regions than O-COMs is Orion KL, where the N-COMs are associated with the hot core while O-COMs are predominantly found toward the compact ridge \citep[e.g.,][]{Blake1987,Friedel2008}. The total abundance of N-COMs in the hot core is more than an order of magnitude higher than in the colder compact ridge \citep{Crockett2014}. If what we are seeing here is the results of carbon grain sublimation, this process may thus indeed result in observable enhancements of N-COMs.

A similar spatial differentiation was also observed toward the high-mass star-forming region W3, where N-COMs are found only toward the W3(H2O) core while O-COMs are present in both the W3(H2O) and W3(OH) regions \citep{Wyrowski1999}, and G34.26+0.15, where emission from N- and O-COMs peak at different positions within the core \citep{Mookerjea2007}. Recent high-resolution observations ($\leq0.5^{\prime\prime}$) reveal differences in AFGL 2591 VLA 3 \citep{Jimenez-Serra2012,Gieser2019}, AFGL 4176 \citep{Bogelund2019a}, G35.20 core B \citep{Allen2017}, and G328.2551-0.5321 \citep{Csengeri2019}. In these high-mass sources, the N-COMs are found to peak on source, while the O-COMs peak offset from the central protostar. In addition, \citet{Fayolle2015} find that the N-COMs are generally concentrated toward the source centers in the three massive young stellar objects NGC 7538 IRS9, W3 IRS5 and AFGL490, while O-COM emission is more extended. 

A hint of more compact emission from N-COMs than O-COMs is also observed for the disk around the outbursting star V883-Ori \citep{Lee2019}. This result is tentative, because due to the vertical temperature structure of disks and the dust being optically thick in the inner $\sim$40 au in this system \citep{Cieza2016} we may be seeing effects in the disk surface layers rather than the inner-disk midplane. 

However, the spatial differentation between N-COMs and O-COMs is not always black and white. For example, the O-COMs ethylene glycol ((CH$_2$OH)$_2$) and acetic acid (CH$_3$COOH) have a spatial distribution in Orion KL different from most other O-bearing molecules \citep{Brouillet2015,Favre2017} and the overall distribution of acetone ((CH$_3$)$_2$CO) is similar to that of ethyl cyanide (C$_2$H$_5$CN) \citep{Peng2013}.  %The latter could be due to the involvement of N-bearing species in the formation or destruction of acetone \citep{Chen2011}, illustrating that chemical relationships between species could cloud the picture of spatial separation between N-COMs and O-COMs.   

Another hint of a difference between N-COMs and O-COMs is found when comparing molecular abundances. Abundances of O-COMs are often found to be correlated, while N-COMs either tend to show stronger correlations with other N-COMs rather than O-COMs, or show no correlation with any other molecule \citep[e.g.,][]{Bisschop2007,Bergner2017,Suzuki2018}. This is also not uniform as, for example, C$_2$H$_3$CN and C$_2$H$_5$CN were found to be correlated with CH$_3$OCH$_3$ by \citet{Fontani2007}, and \citet{Belloche2020} found a strong correlation between CH$_3$OH and CH$_3$CN. However, these results should be regarded with caution, as a correlation across sources does not necessarily reflect an actual chemical relationship but may also reflect underlying physical differences (e.g., temperatures, column densities or simply variations in the reference species).

%............. Figure: Excitation temperatures  ..................................
\begin{figure}[ht!]
\centering
\includegraphics[trim={0.65cm 5.15cm 8.1cm 0.8cm},clip]{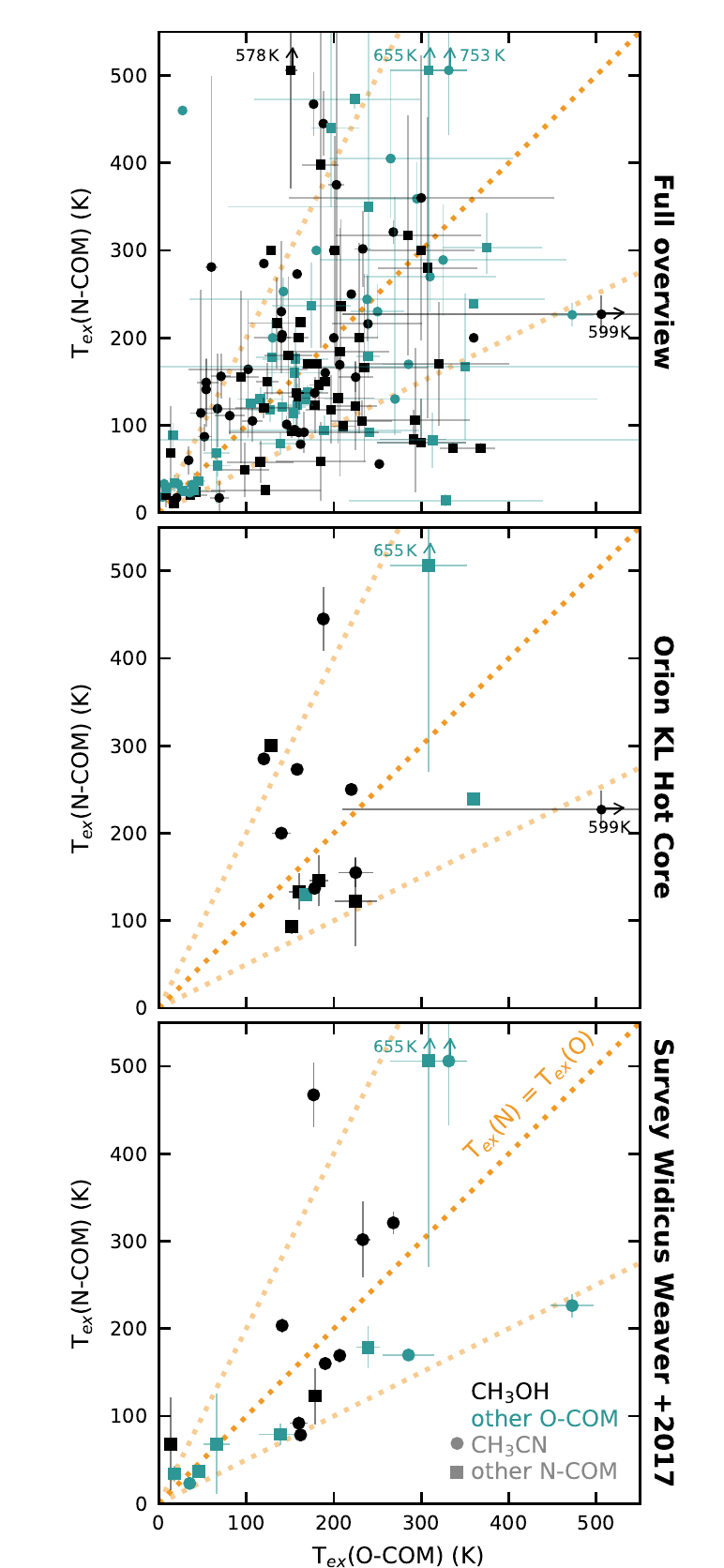}
\caption{Highest excitation temperature among all nitrogen-bearing COMs versus that among all oxygen-bearing COMs reported in a study for a protostellar source. The \textit{top panel} provides an overview of the literature, while the \textit{middle panel} only shows results for Orion KL and the \textit{bottom panel} only shows results from the survey by \citet{WidicusWeaver2017}. References are listed in Table~\ref{tab:Tex}. The shape of the symbols indicate whether the highest excitation temperature for the N-COMs is for CH$_3$CN (circle) or for another species (square). The color of the symbols indicate whether the highest excitation temperature for the O-COMs is for CH$_3$OH (black) or for another species (teal). The dashed lines mark where the excitation temperature for N-COMs is twice that, equal to, and half of the O-COMs.}
\label{fig:Tex}
\end{figure}
%....................................................................................

\subsection{Excitation temperature of N-COMs and O-COMs}

Differences in excitation temperatures of N-COMs compared to O-COMs have been observed for Orion KL, with the N-COMs tracing hotter gas ($\sim$300~K) than the O-COMs \citep{Crockett2015}. The latter are consistent with sublimation alongside water around $\sim$100 K. A similar picture emerges for Sgr B2(N2), although the temperature differences are smaller \citep{Belloche2016}. On the other hand, O-COMs toward IRAS 16293--2422 B can be divided in two groups based on their excitation temperature ($\sim$125 K versus $\sim$300 K; \citealt{Jorgensen2018}), consistent with different binding energies. N-COMs generally have excitation temperatures of 100--150 K \citep{Calcutt2018a,Calcutt2018b}, but a high-temperature component may be hidden inside the unresolved soot line or by the optically thick dust.

To assess whether a difference in excitation temperature is a widespread phenomenon, we compiled an overview of the existing literature that reports excitation temperatures for at least one N-COM and one O-COM. Because not all studies observe the same molecules, Fig.~\ref{fig:Tex} (top panel, and see also Table~\ref{tab:Tex}) shows the highest reported excitation temperature for N-COMs versus that of O-COMs for each source in a particular study. Molecules with less than 5 atoms and molecules containing both nitrogen and oxygen, like HNCO and NH$_2$CHO, are excluded. While there clearly are a large number of cases where a higher excitation temperature is reported for N-COMs than for O-COMs (64 out of 144 entries; 44\%), this is certainly not always the case. Possible explanations will be discussed in Sect.~\ref{sec:Discussion}.

% ======================================================================
% DISCUSSION
% ======================================================================

\section{Discussion and outlook} \label{sec:Discussion}

The spatial differentiation between N-COMs and O-COMs has been a longstanding problem \citep{Blake1987}. Initially, this was attributed to different ice compositions, because the relative amount of ammonia (NH$_3$) ice had a large impact on the CH$_3$CN/CH$_3$OH ratio in the chemical models from \citet{Charnley1992} and \citet{Rodgers2001}. However, more recent models do not predict a correlation between the CH$_3$CN/CH$_3$OH gas ratio and the NH$_3$/CH$_3$OH ice ratio \citep{Garrod2013}, consistent with observations of both low- and high-mass protostars \citep{Fayolle2015,Bergner2017}. Another explanation given in the literature is based on a difference in temperature between regions in combination with a different warm-up timescale, hence evolutionary stage \citep[e.g.,][]{Caselli1993,Garrod2008,Allen2018}.

Here, we suggest the thermal destruction of carbon grains inside the soot line ($\sim$ 300 K) as the underlying reason for the differences in distribution between N-COMs and O-COMs: regions rich in N-COMs are currently heated to temperatures above $\sim$ 300 K, or have been in the past. While a different thermal history and evolutionary stage may explain the spatial differentiation and differences in excitation temperature, only carbon-grain sublimation can simultaneously explain the low carbon and nitrogen content of the Earth and CI chondrites, and the carbon-depleted pollution seen in white dwarf atmospheres. 

As outlined in Sect.~\ref{sec:Evidence} and visualized in Fig.~\ref{fig:Tex}, a coherent picture indicating that carbon-grain sublimation is a common phenomenon during star formation does not yet exist. A complicating factor that could contribute to the non-uniformity in observed spatial distributions and abundance correlations, is that carbonaceous material could contain a small amount of oxygen.  For example, about 25\% of elemental oxygen is unaccounted for in the interstellar medium \citep[e.g.,][]{Whittet2010, Poteet2015}. There is some insight here from meteorites, as insoluble organic matter contains 22 O atoms for every 100 C atoms \citep{Remusat2014}, which is roughly consistent with interstellar inferences. Thus, there is likely some oxygen-bearing organic material released as carbon grains are ablated. 

Figure~\ref{fig:Tex} displays a wide spread in excitation temperatures derived for both N-COMs and O-COMs (from a few K up to $>$ 500 K). In addition, there is no clear correlation between the highest temperature for N-COMs and that for O-COMs, as N-COMs can be more than twice as hot as O-COMs but also more than twice as cold. The spread in excitation temperatures could be due to the heterogeneity of observations and analysis methods (see Fig.~\ref{fig:Tex}, middle panel, for an overview of studies toward Orion KL). Besides different spatial resolutions with respect to the size of the soot line, a relevant difference between studies is the covered wavelength range, as this determines which molecules are observed, the number of lines per molecule and the upper level energies of these lines. The latter can complicate conclusions about the spatial extent of a molecule, as even typical hot-core molecules, such as CH$_3$OH and CH$_3$CN, can be observed in the outer envelope when observing low-energy transitions \citep[e.g.,][]{Oberg2013}. In addition, these molecules often require both a hot and cold component to explain observations, and this hot component could be missed when only lines with low upper level energies are targeted. 

Furthermore, differences could result from different analysis techniques. In general, excitation temperatures are derived from rotational diagrams or from fitting the observed spectrum, but different assumptions about beam dilution and/or optical depth could give significantly different results (see e.g., \citealt{Gibb2000}). Finally, rotational temperatures can be higher than the kinetic temperature due to optical depth effects and/or infrared pumping \citep[see e.g.][]{Churchwell1986} and for large-scale single-dish studies the critical density and dipole moment of a species may be more important for the excitation than the temperature. Equal or lower excitation temperatures for N-COMs therefore do not necessarily indicate that carbon-grain sublimation is not occurring. 

However, also systematic surveys do not present a uniform picture (see Fig.~\ref{fig:Tex}, bottom panel, for results from the survey by \citealt{WidicusWeaver2017}). The spread could illustrate that the soot line is not resolved in all sources, but sources with higher N-COM excitation temperatures span the entire luminosity and distance range among the high-mass sources in the sample. Another explanation could therefore be that a clear signature of carbon-grain sublimation may not be present for all sources. The episodic nature of the protostellar accretion could erase the spatial difference between N-COMs and O-COMs as well as differences in their excitation temperature (see Fig.~\ref{fig:Cartoon}). The fraction of sources in the post-burst phase (with the O-COMs spatially more extended than the N-COMs) is equal to the freeze-out timescale divided by the burst interval. The O-COMs have desorption temperatures similar to water, so their freeze-out timescale is $\sim$1000 years \citep{Visser2015}. Several studies estimate that the burst interval is 20,000-50,000 years \citep[e.g.,][]{Scholz2013,Jorgensen2015}, while \citet{Hsieh2019} derive an interval of $\sim$2400 years during the Class 0 phase. This means that we would expect 2-5\% up to 40\% of protostars in the post-burst phase. These numbers are not inconsistent with the number of studies finding higher excitation temperatures for N-COMs than for O-COMs in Fig.~\ref{fig:Tex}. A caveat here is, however, that it is unclear whether burst statistics for low-mass sources also apply for high-mass sources, which make up the majority (80\%) of Fig.~\ref{fig:Tex}.

To unambigously establish the presence of carbon-grain sublimation, large systematic surveys are thus required that spatially probe the soot line as well as cover a large number of lines with a range of upper level energies of many different molecular species. Chemical models incorporating the thermal destruction of carbon grains (such as done for protoplanetary disks by \citealt{Wei2019}) could help indicate what species are good indicators. The best candidates to establish the occurrence of carbon grain sublimation are sources with a clear spatial differentiation between N- and O-COMs (as listed in Sect.~\ref{sec:ObsSpatDis}). A deep spectral survey and detailed analysis (e.g., of warm hydrocarbons) could reveal whether the observed distributions are consistent with suggested accretion shocks at the disk-envelope interface or with sublimation of carbon grains. Although the study of low-mass sources may be more appropriate to understand the formation of the Solar System and its analogs, the larger $>$300~K region makes high-mass sources the more accessible astrochemical laboratories.

If carbon grain sublimation is happening, it would mean that there is an unexplored top-down component to protostellar chemistry, that starts from the destruction of larger structures instead of from atoms. The implications for astrochemistry, the origin of organic molecules and carbon delivery to terrestrial worlds then needs to be re-explored. 

% ======================================================================
% ACKNOWLEDGMENTS
% ======================================================================

\acknowledgments 

We would like to thank the referee and Ewine van Dishoeck for positive feedback that has improved this paper. M.L.R.H. acknowledges support from the Michigan Society of Fellows. E.A.B. acknowledges funding from NSF grant AST 1907653 and NASA grant XRP 80NSSC20K0259. J.K.J. acknowledges support by the European Research Council (ERC) under the European Union's Horizon 2020 research and innovation programme through ERC Consolidator Grant ``S4F'' (grant agreement No~646908). G.A.B. acknowledges support from the NASA XRP (NNX16AB48G) and Astrobiology (NNX15AT33A) programs.

% ......... Table: Excitation temperatures ..............................

\begin{longrotatetable}
\begin{deluxetable*}{l l c c c c c c c l}
\tablecaption{Overview of literature presenting excitation temperatures for both N- and O-COMs in protostars.
\label{tab:Tex}}
\tablewidth{700pt}
\tabletypesize{\scriptsize}
\tablehead{
\colhead{Source name} \vspace{-0.2cm}& \colhead{Source type} & \colhead{$T_{\rm{ex}}$(N-COM)\tablenotemark{a}} & \colhead{$T_{\rm{ex}}$(O-COM)\tablenotemark{b}} & \colhead{$T_{\rm{ex}}$(CH$_3$CN)\tablenotemark{c}} & \colhead{$T_{\rm{ex}}$(CH$_3$OH)\tablenotemark{d}} & \colhead{Sign of} & \colhead{Telescope} & \colhead{Analysis} & \colhead{Reference}  \\ 
\colhead{} \vspace{-0.5cm}& \colhead{} & \colhead{(K)} & \colhead{(K)} & \colhead{(K)} & \colhead{(K)} & \colhead{CGS\tablenotemark{e}} & \colhead{} & \colhead{method\tablenotemark{f}} \\ 
} 
\startdata 
AFGL 2591 VLA 3 & High mass  & 150 $\pm$ 20 & 124 $\pm$ 12 & n/a & \nodata & yes & SMA & RD & \citet{Jimenez-Serra2012} \\ 
\nodata & \nodata & 218 $\pm$ 2 & 162 $\pm$ 2 & 199 $\pm$ 1 & \nodata & yes & IRAM 30m + NOEMA & M & \citet{Gieser2019} \\ 
AFGL 4176  & High mass & 270 $\pm$ 40 & 310 $\pm$ 75 & \nodata & 120 $\pm$ 15 & \nodata & ALMA & M & \citet{Bogelund2019a} \\ 
AFGL 490  & High mass & 164 $\pm$ 78 & 102 $\pm$ 67 & \nodata & \nodata & yes & IRAM 30m + SMA & RD & \citet{Fayolle2015} \\ 
B1-a  & Low mass & 33 $\pm$ 9 & 22 $\pm$ 5 & \nodata & n/a & yes & IRAM 30m & RD & \citet{Bergner2017} \\ 
B1-c  & Low mass & 18 $\pm$ 2 & 4 $\pm$ 1 & \nodata & n/a & yes & IRAM 30m & RD & \citet{Bergner2017} \\ 
B5 IRS 1  & Low mass & 11 $\pm$ 2 & 17 $\pm$ 2 & n/a & \nodata & \nodata & IRAM 30m & RD & \citet{Oberg2014,Bergner2017} \\ 
Cep E-A  & Int. mass  & 32 $\pm$ 4 & 37 $\pm$ 1 & \nodata & 27 $\pm$ 5 & \nodata & IRAM 30m + NOEMA & RD & \citet{Ospina-Zamudio2018} \\ 
DR21(OH)  & High mass & 55 & 252 & \nodata & \nodata & \nodata & OSO & RD & \citet{Kalenskii2010a} \\ 
\nodata & \nodata & 79 $\pm$ 11 & 138 $\pm$ 24 & 56 $\pm$ 2 & 92 $\pm$ 2 & \nodata & CSO & M & \citet{WidicusWeaver2017} \\ 
G9.62+0.19  & High mass & $119^{62}_{43}$ & $67^{57}_{23}$ & \nodata & \nodata & yes & JCMT & M & \citet{Hatchell1998} \\ 
G10.47+0.03  & High mass & $87^{14}_{16}$ & $52^{8}_{6}$ & \nodata & \nodata & yes & JCMT & M & \citet{Hatchell1998} \\ 
\nodata & \nodata & 176 $\pm$ 35 & 156 $\pm$ 37 & n/a & n/a & yes & IRAM 30m & RD & \citet{Fontani2007} \\ 
\nodata & \nodata & 753 $\pm$ 72 & 331 $\pm$ 20 & \nodata & 258 $\pm$ 1 & yes & CSO & M & \citet{WidicusWeaver2017} \\ 
\nodata & \nodata & 398 $\pm$ 209 & 185 $\pm$ 20 & n/a & \nodata & yes & Nobeyama & RD & \citet{Suzuki2018} \\ 
\nodata & \nodata & 217 $\pm$ 51 & 135 $\pm$ 32 & n/a & \nodata & yes & Nobeyama & RD & \citet{Suzuki2018} \\ 
\nodata & \nodata & 166 $\pm$ 106 & 235 $\pm$ 71 & n/a & \nodata & \nodata & ALMA & RD & \citet{Suzuki2019} \\ 
G10.62--0.38  & High mass & 89 $\pm$ 13 & 16 $\pm$ 1 & n/a & n/a & yes & IRAM 30m & RD & \citet{Fontani2007} \\ 
\nodata & \nodata & 178 $\pm$ 42 & 129 $\pm$ 17 & 165 $\pm$ 23 & n/a & yes & SMA & M & \citet{Wong2018} \\ 
G12.21--0.10  & High mass & 78 $\pm$ 4 & 161 $\pm$ 3 & \nodata & \nodata & \nodata & CSO & M & \citet{WidicusWeaver2017} \\ 
G12.89+0.49  & High mass & 24 & $42^{33}_{15}$ & n/a & \nodata & \nodata & Nobeyama 45m + ASTE & RD & \citet{Taniguchi2018} \\ 
G12.91--0.26  & High mass & 122 $\pm$ 31 & 178 $\pm$ 4 & 116 $\pm$ 5 & \nodata & \nodata & CSO & M & \citet{WidicusWeaver2017} \\ 
G16.86--2.16  & High mass & 20 & $36^{18}_{9}$ & n/a & \nodata & \nodata & Nobeyama 45m + ASTE & RD & \citet{Taniguchi2018} \\ 
G19.61-0.23  & High mass & 123 $\pm$ 24 & 158 $\pm$ 17 & n/a & n/a & \nodata & IRAM 30m & RD & \citet{Fontani2007} \\ 
\nodata & \nodata & 578 $\pm$ 134 & 151 $\pm$ 6 & n/a & \nodata & yes & SMA & RD & \citet{Qin2010} \\ 
\nodata & \nodata & 467 $\pm$ 35 & 176 $\pm$ 2 & \nodata & \nodata & yes & CSO & M & \citet{WidicusWeaver2017} \\ 
G24.33+00.11 MM1 & High mass & 301 $\pm$ 42 & 233 $\pm$ 8 & \nodata & \nodata & yes & CSO & M & \citet{WidicusWeaver2017} \\ 
G24.78+0.08  & High mass & $99^{5}_{0}$ & 211 $\pm$ 13 & n/a & \nodata & \nodata & JCMT + IRAM 30m & RD & \citet{Bisschop2007} \\ 
\nodata & \nodata & 203 $\pm$ 7 & 140 $\pm$ 1 & \nodata & \nodata & yes & CSO & M & \citet{WidicusWeaver2017} \\ 
G28.28--0.36  & High mass & 13 & $18^{5}_{3}$ & n/a & \nodata & \nodata & Nobeyama 45m + ASTE & RD & \citet{Taniguchi2018} \\ 
G29.96--0.02  & High mass & $114^{140}_{80}$ & $48^{17}_{11}$ & \nodata & \nodata & yes & JCMT & M & \citet{Hatchell1998} \\ 
\nodata & \nodata & 121 $\pm$ 17 & 141 $\pm$ 26 & n/a & n/a & \nodata & IRAM 30m & RD & \citet{Fontani2007} \\ 
G31.41+0.31  & High mass & $149^{27}_{36}$ & $54^{13}_{10}$ & \nodata & \nodata & yes & JCMT & M & \citet{Hatchell1998} \\ 
\nodata & \nodata & 118 $\pm$ 13 & 127 $\pm$ 25 & n/a & n/a & \nodata & IRAM 30m & RD & \citet{Fontani2007} \\ 
\nodata & \nodata & $105^{23}_{0}$ & 232 $\pm$ 34 & n/a & \nodata & \nodata & JCMT & RD & \citet{Isokoski2013} \\ 
\nodata & \nodata & $80^{50}_{0}$ & 300 $\pm$ 50 & n/a & 200 $\pm$ 50 & \nodata & JCMT & M & \citet{Isokoski2013} \\ 
\nodata & \nodata & 320 $\pm$ 12 & 268 $\pm$ 4 & \nodata & \nodata & yes & CSO & M & \citet{WidicusWeaver2017} \\ 
\nodata & \nodata & 59 $\pm$ 44 & 185 $\pm$ 51 & n/a & \nodata & \nodata & Nobeyama & RD & \citet{Suzuki2018} \\ 
\nodata & \nodata & 184 $\pm$ 141 & 207 $\pm$ 55 & n/a & \nodata & \nodata & ALMA & RD & \citet{Suzuki2019} \\ 
G34.26+0.15  & High mass & 74 $\pm$ 5 & 368 $\pm$ 16 & n/a & \nodata & \nodata & JCMT & RD & \citet{MacDonald1996} \\ 
\nodata & \nodata & 74 $\pm$ 5 & 336 $\pm$ 14 & n/a & \nodata & \nodata & JCMT & RD & \citet{MacDonald1996} \\ 
\nodata & \nodata & $141^{34}_{42}$ & $54^{3}_{2}$ & \nodata & \nodata & yes & JCMT & M & \citet{Hatchell1998} \\ 
\nodata & \nodata & 130 $\pm$ 13 & 116 $\pm$ 25 & n/a & n/a & yes & IRAM 30m & RD & \citet{Fontani2007} \\ 
\nodata & \nodata & 300 $\pm$ 130 & 201 $\pm$ 4 & 298 $\pm$ 8 & \nodata & yes & SMA & RD & \citet{Fu2016} \\ 
\nodata & \nodata & 160 $\pm$ 3 & 190 $\pm$ 1 & \nodata & \nodata & \nodata & CSO & M & \citet{WidicusWeaver2017} \\ 
\nodata & \nodata & 167 $\pm$ 83 & 350 $\pm$ 412 & n/a & 196 $\pm$ 42 & \nodata & Nobeyama & RD & \citet{Suzuki2018} \\ 
\nodata & \nodata & 49 $\pm$ 30 & 98 $\pm$ 27 & n/a & \nodata & \nodata & Nobeyama & RD & \citet{Suzuki2018} \\ 
G35.03+0.35 A & High mass & $216^{54}_{19}$ & $239^{61}_{40}$ & \nodata & \nodata & \nodata & ALMA  & M & \citet{Allen2017} \\ 
G35.20--0.74 N A & High mass & $359^{41}_{54}$ & $295^{5}_{92}$ & \nodata & $227^{73}_{18}$ & yes & ALMA & M & \citet{Allen2017} \\ 
G35.20--0.74 N B3 & High mass & $473^{4}_{10}$ & $224^{74}_{114}$ & $213^{27}_{47}$ & 189 $\pm$ 14 & yes & ALMA & M & \citet{Allen2017} \\ 
G45.47+0.05  & High mass & 68 $\pm$ 52 & 13 $\pm$ 0 & n/a & \nodata & yes & CSO & M & \citet{WidicusWeaver2017} \\ 
G75.78+0.34  & High mass & 91 $\pm$ 7 & 159 $\pm$ 9 & \nodata & \nodata & \nodata & CSO & M & \citet{WidicusWeaver2017} \\ 
G327.3--0.60  & High mass & $303^{39}_{29}$ & $375^{62}_{50}$ & $282^{125}_{65}$ & $180^{14}_{11}$ & \nodata & Herschel HIFI & RD & \citet{Gibb2000} \\ 
\nodata & \nodata & $138^{7}_{9}$ & $170^{16}_{18}$ & n/a & 102 $\pm$ 10 & \nodata & Herschel HIFI & RD & \citet{Gibb2000} \\ 
\nodata & \nodata & 135 $\pm$ 5 & $162^{11}_{9}$ & n/a & $118^{6}_{8}$ & \nodata & Herschel HIFI & RD & \citet{Gibb2000} \\ 
G331.512--0.103  & High mass & 156 $\pm$ 25 & 71 $\pm$ 10 & \nodata & \nodata & yes & APEX & RD & \citet{Mendoza2018} \\ 
IRAS 03235+3004  & Low mass & 13 $\pm$ 2 & 18 $\pm$ 4 & n/a & \nodata & \nodata & IRAM 30m & RD & \citet{Oberg2014,Bergner2017} \\ 
IRAS 16293--2422  & Low mass & 54 $\pm$ 11 & 67 $\pm$ 14 & n/a & n/a & \nodata & IRAM 30m & RD & \citet{Cazaux2003} \\ 
IRAS 16293--2422 A & Low mass & 160 $\pm$ 20 & 155 $\pm$ 31 & 120 $\pm$ 10 & 130 $\pm$ 26 & yes & ALMA & M & \citet{Calcutt2018a,Manigand2020} \\ 
IRAS 16293--2422 B & Low mass & 300 & 300 $\pm$ 60 & n/a & \nodata & \nodata & ALMA & M & \citet{Coutens2018,Jorgensen2018} \\ 
IRAS 18089--1732  & High mass & $84^{33}_{0}$ & 291 $\pm$ 37 & n/a & \nodata & \nodata & JCMT & RD & \citet{Isokoski2013} \\ 
\nodata & \nodata & $80^{50}_{0}$ & 300 $\pm$ 50 & n/a & \nodata & \nodata & JCMT & M & \citet{Isokoski2013} \\ 
IRAS 20126+4104  & High mass & 230 $\pm$ 30 & 250 $\pm$ 30 & \nodata & n/a & \nodata & PdBI & M & \citet{Palau2017} \\ 
IRAS 23238+7401  & Low mass & 33 $\pm$ 6 & 6 $\pm$ 2 & \nodata & n/a & yes & IRAM 30m & RD & \citet{Bergner2017} \\ 
IRDC-C9  Main & High mass & 460 & 27 $\pm$ 5 & \nodata & n/a & yes & ALMA & RD & \citet{Beaklini2020} \\ 
L1448-C  & Low mass & 105 $\pm$ 23 & 107 $\pm$ 11 & \nodata & \nodata & \nodata & NOEMA & RD & \citet{Belloche2020} \\ 
L1489 IRS  & Low mass & 20 $\pm$ 13 & 8 $\pm$ 4 & n/a & \nodata & yes & IRAM 30m & RD & \citet{Oberg2014,Bergner2017} \\ 
L1527  & Low mass & 25 $\pm$ 4 & 27 $\pm$ 23 & 21 $\pm$ 2 & 13 $\pm$ 8 & \nodata & NRO 45m & RD & \citet{Yoshida2019} \\ 
L483  & Low mass & 28 $\pm$ 3 & 7 $\pm$ 13 & n/a & n/a & yes & IRAM 30m & M & \citet{Agundez2019} \\ 
NGC1333-IRAS2 A & Low mass & 289 $\pm$ 63 & 325 $\pm$ 140 & \nodata & 179 $\pm$ 62 & \nodata & PdBI & RD & \citet{Taquet2015} \\ 
\nodata & \nodata & $130^{230}_{40}$ & $270^{230}_{80}$ & \nodata & 140 $\pm$ 20 & \nodata & PdBI & RD & \citet{Taquet2015} \\ 
NGC1333-IRAS2 A1 & Low mass & 200 $\pm$ 108 & 229 $\pm$ 21 & 161 $\pm$ 17 & \nodata & \nodata & NOEMA & RD & \citet{Belloche2020} \\ 
NGC1333-IRAS4 A & Low mass & 360 $\pm$ 162 & 300 $\pm$ 151 & \nodata & \nodata & yes & PdBI & RD & \citet{Taquet2015} \\ 
\nodata & \nodata & $200^{110}_{40}$ & 140 $\pm$ 30 & \nodata & \nodata & yes & PdBI & RD & \citet{Taquet2015} \\ 
NGC1333-IRAS4 A2 & Low mass & 280 $\pm$ 171 & 307 $\pm$ 56 & 142 $\pm$ 13 & \nodata & \nodata & NOEMA & RD & \citet{Belloche2020} \\ 
NGC1333-IRAS4 B & Low mass & 14 $\pm$ 6 & 328 $\pm$ 110 & n/a & 305 $\pm$ 52 & \nodata & NOEMA & RD & \citet{Belloche2020} \\ 
NGC 2264 CMM3  & High mass & 25 $\pm$ 4 & 122 $\pm$ 62 & 9 $\pm$ 4 & \nodata & \nodata & Nobeyama 45m / ASTE & RD & \citet{Watanabe2015} \\ 
NGC 6334-29  & High mass & 169 $\pm$ 4 & 285 $\pm$ 28 & \nodata & 127 $\pm$ 1 & \nodata & CSO & M & \citet{WidicusWeaver2017} \\ 
NGC 6334-38  & High mass & 36 $\pm$ 2 & 45 $\pm$ 6 & 33 $\pm$ 1 & 21 $\pm$ 0 & \nodata & CSO & M & \citet{WidicusWeaver2017} \\ 
NGC 6334-43  & High mass & 33 $\pm$ 8 & 17 $\pm$ 4 & 17 $\pm$ 2 & 16 $\pm$ 0 & yes & CSO & M & \citet{WidicusWeaver2017} \\ 
NGC 6334-I(N)  & High mass & 22 $\pm$ 1 & 35 $\pm$ 3 & \nodata & 20 $\pm$ 0 & \nodata & CSO & M & \citet{WidicusWeaver2017} \\ 
NGC 6334 IRS1  & High mass & $92^{3}_{0}$ & 241 $\pm$ 35 & n/a & 178 $\pm$ 10 & \nodata & JCMT + IRAM 30m & RD & \citet{Bisschop2007} \\ 
\nodata & \nodata & 58 $\pm$ 23 & 116 $\pm$ 37 & n/a & \nodata & \nodata & Nobeyama & RD & \citet{Suzuki2018} \\ 
NGC 6334 IRS1 F & High mass & 83 $\pm$ 31 & 313 $\pm$ 347 & n/a & 201 $\pm$ 24 & \nodata & Nobeyama & RD & \citet{Suzuki2018} \\ 
NGC 6334I MM1 & High mass & 120 $\pm$ 11 & 120 $\pm$ 32 & n/a & \nodata & \nodata & ALMA & RD & \citet{Suzuki2019} \\ 
NGC 6334I MM2 & High mass & 137 $\pm$ 46 & 157 $\pm$ 36 & n/a & \nodata & \nodata & ALMA & RD & \citet{Suzuki2019} \\ 
NGC 7129 FIRS 2  & Int. mass & $405^{100}_{67}$ & $265^{139}_{69}$ & \nodata & $157^{15}_{13}$ & yes & PdBI & RD & \citet{Fuente2014} \\ 
\nodata & \nodata & 405 & 265 & \nodata & 238 & yes & PdBI & M & \citet{Fuente2014} \\ 
NGC 7538 14.37 & High mass & 68 $\pm$ 56 & 66 $\pm$ 14 & n/a & n/a & yes & CSO & M & \citet{WidicusWeaver2017} \\ 
NGC 7538 IRS9  & High mass & 281 $\pm$ 218 & 60 $\pm$ 6 & \nodata & \nodata & yes & IRAM 30m + SMA & RD & \citet{Oberg2013} \\ 
\nodata & \nodata & 60 $\pm$ 15 & 34 $\pm$ 2 & \nodata & \nodata & yes &  & RD & \citet{Oberg2013} \\ 
\nodata & \nodata & 111 $\pm$ 20 & 81 $\pm$ 16 & \nodata & \nodata & yes & IRAM 30m + SMA & RD & \citet{Fayolle2015} \\ 
Orion KL  & High mass & 200 & 140 $\pm$ 10 & \nodata & \nodata & yes & OSO & RD & \citet{Johansson1984} \\ 
\nodata & \nodata & 285 & 120 & \nodata & \nodata & yes & OVRO & RD & \citet{Sutton1985} \\ 
\nodata & \nodata & 445 $\pm$ 36 & 188 $\pm$ 3 & \nodata & \nodata & yes & CSO & RD & \citet{Schilke1997} \\ 
\nodata & \nodata & 273 & 158 & \nodata & \nodata & yes & TRAO 14m & RD & \citet{Lee2001} \\ 
\nodata & \nodata & 227 $\pm$ 21 & 599 $\pm$ 295 & \nodata & \nodata & \nodata & JCMT & RD & \citet{White2003} \\ 
\nodata & \nodata & 250 & 220 & \nodata & \nodata & yes & CSO & M & \citet{Comito2005} \\ 
\nodata & \nodata & 655 $\pm$ 235 & 308 $\pm$ 42 & 229 $\pm$ 4 & 194 $\pm$ 2 & yes & CSO & M & \citet{WidicusWeaver2017} \\ 
\nodata & \nodata & 146 $\pm$ 28 & 183 $\pm$ 10 & n/a & \nodata & \nodata & Nobeyama & RD & \citet{Suzuki2018} \\ 
\nodata & \nodata & 122 $\pm$ 50 & 225 $\pm$ 23 & n/a & \nodata & \nodata & Nobeyama & RD & \citet{Suzuki2018} \\ 
Orion KL hot core & High mass & 133 $\pm$ 20 & 160 $\pm$ 10 & n/a & \nodata & \nodata & Nobeyama & RD & \citet{Ohishi1986} \\ 
\nodata & \nodata & 130 & 168 & n/a & n/a & \nodata & JCMT & RD  & \citet{Sutton1995} \\ 
\nodata & \nodata & 93 $\pm$ 7 & 152 & n/a & \nodata & \nodata & NRO, NRAO & RD & \citet{Ikeda2001} \\ 
\nodata & \nodata & 239 & 360 & n/a & 303 & \nodata & CSO & RD & \citet{Schilke2001} \\ 
\nodata & \nodata & 137 & 178 & \nodata & \nodata & \nodata & Odin & RD & \citet{Persson2007} \\ 
\nodata & \nodata & 300 & 128 & 260 & \nodata & yes & Herschel HIFI & M & \citet{Crockett2014} \\ 
\nodata & \nodata & 155 $\pm$ 16 & 225 $\pm$ 19 & \nodata & \nodata & \nodata & IRAM 30m + SMA & RD & \citet{Feng2015} \\ 
Orion KL compact ridge & High mass & 101 $\pm$ 9 & 146 $\pm$ 3 & \nodata & \nodata & \nodata & OVRO & RD & \citet{Blake1987} \\ 
\nodata & \nodata & 125 & 105 & n/a & n/a & yes & JCMT & RD  & \citet{Sutton1995} \\ 
\nodata & \nodata & 230 & 140 & \nodata & \nodata & yes & Herschel HIFI & M & \citet{Crockett2014} \\ 
SerpS-MM18a  & Low mass & 244 $\pm$ 28 & 238 $\pm$ 202 & \nodata & 154 $\pm$ 22 & yes & NOEMA & RD & \citet{Belloche2020} \\ 
SerpS-MM18b  & Low mass & 17 $\pm$ 18 & 69 $\pm$ 10 & \nodata & \nodata & \nodata & NOEMA & RD & \citet{Belloche2020} \\ 
Sgr B2(N)  & High mass & $440^{190}_{90}$ & $197^{31}_{22}$ & $400^{104}_{86}$ & n/a & yes & SEST & RD & \citet{Nummelin2000} \\ 
\nodata & \nodata & 180 $\pm$ 89 & 148 $\pm$ 26 & n/a & \nodata & yes & NRO & RD & \citet{Ikeda2001} \\ 
\nodata & \nodata & 200 & 130 & \nodata & n/a & yes & IRAM 30m & M & \citet{Belloche2009} \\ 
\nodata & \nodata & 200 & 200 & \nodata & \nodata & \nodata & IRAM 30m & M & \citet{Belloche2013} \\ 
\nodata & \nodata & 300 & 180 & \nodata & 170 & yes & Herschel HIFI & M & \citet{Neill2014} \\ 
Sgr B2(N) N2 & High mass & 253 $\pm$ 15 & 142 $\pm$ 4 & \nodata & n/a & yes & ALMA & RD & \citet{Belloche2016} \\ 
\nodata & \nodata & 200 & 160 & 170 & \nodata & yes & ALMA & M & \citet{Belloche2016} \\ 
Sgr B2(N) N3 & High mass & 170 & 170 & 145 & \nodata & \nodata & ALMA & M & \citet{Bonfand2017} \\ 
Sgr B2(N) N4 & High mass & 150 & 190 & 145 & \nodata & \nodata & ALMA & M & \citet{Bonfand2017} \\ 
Sgr B2(N) N5 & High mass & 170 & 180 & 145 & \nodata & \nodata & ALMA & M & \citet{Bonfand2017} \\ 
Sgr B2(M)  & High mass & 300 & 200 & \nodata & \nodata & yes & IRAM 30m & M & \citet{Belloche2013} \\ 
\nodata & \nodata & $350^{510}_{140}$ & $240^{0}_{160}$ & n/a & 150 & yes & SEST & RD & \citet{Nummelin2000} \\ 
Sgr B2(NW)  & High mass & $26^{19}_{8}$ & $40^{13}_{8}$ & n/a & $16^{4}_{2}$ & \nodata & SEST & RD & \citet{Nummelin2000} \\ 
SVS 4-5  & Low mass & 17 $\pm$ 2 & 20 $\pm$ 0 & \nodata & \nodata & \nodata & IRAM 30m & RD & \citet{Oberg2014,Bergner2017} \\ 
SVS13-A  & Low mass & 317 $\pm$ 136 & 285 $\pm$ 82 & 222 $\pm$ 49 & \nodata & yes & NOEMA & RD & \citet{Belloche2020} \\ 
W3(H2O)  & High mass & 375 $\pm$ 210 & 203 $\pm$ 8 & \nodata & \nodata & yes & JCMT   & RD & \citet{Helmich1997} \\ 
\nodata & \nodata & $94^{16}_{0}$ & 189 $\pm$ 108 & n/a & 139 $\pm$ 8 & \nodata & JCMT + IRAM 30m & RD & \citet{Bisschop2007} \\ 
\nodata & \nodata & 200 & 360 & \nodata & \nodata & \nodata & SMA & M & \citet{Qin2015} \\ 
\nodata & \nodata & 178 $\pm$ 23 & 239 $\pm$ 12 & 152 $\pm$ 3 & 151 $\pm$ 1 & \nodata & CSO & M & \citet{WidicusWeaver2017} \\ 
W3(OH)  & High mass & 95 & 155 & \nodata & \nodata & \nodata & SMA & M & \citet{Qin2015} \\ 
W3 IRS5  & High mass & 92 $\pm$ 23 & 166 $\pm$ 85 & \nodata & \nodata & \nodata & IRAM 30m + SMA & RD & \citet{Fayolle2015} \\ 
W43-MM1 core 3 & High mass & 170 $\pm$ 70 & 320 $\pm$ 80 & 130 $\pm$ 60 & \nodata & \nodata & ALMA & RD  & \citet{Molet2019} \\ 
W51  & High mass & 226 $\pm$ 12 & 472 $\pm$ 23 & \nodata & 257 $\pm$ 4 & \nodata & CSO & M & \citet{WidicusWeaver2017} \\ 
\nodata & \nodata & 118 $\pm$ 38 & 197 $\pm$ 28 & n/a & \nodata & \nodata & Nobeyama & RD & \citet{Suzuki2018} \\ 
\nodata & \nodata & 106 $\pm$ 82 & 293 $\pm$ 62 & n/a & \nodata & \nodata & Nobeyama & RD & \citet{Suzuki2018} \\ 
W51 e1/e2 & High mass & 236 $\pm$ 99 & 208 & n/a & \nodata & yes & NRO, NRAO & RD & \citet{Ikeda2001} \\ 
\nodata & \nodata & 237 $\pm$ 48 & 174 $\pm$ 44 & 212 $\pm$ 8 & 143 $\pm$ 10 & yes & OSO & RD & \citet{Kalenskii2010b} \\ 
W51 e2  & High mass & 114 $\pm$ 11 & 154 $\pm$ 8 & n/a & n/a & \nodata & IRAM 30m & RD & \citet{Demyk2008} \\ 
\nodata & \nodata & 155 $\pm$ 98 & 94 $\pm$ 19 & n/a & \nodata & yes & ALMA & RD & \citet{Suzuki2019} \\ 
W51 e8 & High mass & 131 $\pm$ 51 & 205 $\pm$ 40 & n/a & \nodata & \nodata & ALMA & RD & \citet{Suzuki2019} \\ 
W75N  & High mass & 169 $\pm$ 7 & 206 $\pm$ 2 & \nodata & \nodata & \nodata & CSO & M & \citet{WidicusWeaver2017} \\ 
\hline 
144\tablenotemark{g} & & & & & & 64\tablenotemark{h} \\ 
\enddata
%\tablecomments{Table 1 is published in its entirety in the machine-readable format. A portion is shown here for guidance regarding its form and content.}
\tablenotetext{a}{Highest excitation temperature among all N-bearing COMs in the study.}
\vspace{-0.3cm} 
\tablenotetext{b}{Highest excitation temperature among all O-bearing COMs in the study.}
\vspace{-0.3cm} 
\tablenotetext{c}{Excitation temperature of CH$_3$CN (or isotopologue) if this was not the highest temperature for all N-COMs. n/a means no $T_{\rm{ex}}$ is available for CH$_3$CN.}
\vspace{-0.3cm} 
\tablenotetext{d}{Excitation temperature of CH$_3$OH (or isotopologue) if this was not the highest temperature for all O-COMs. n/a means no $T_{\rm{ex}}$ is available for CH$_3$OH.}
\vspace{-0.3cm} 
\tablenotetext{e}{Sign of carbon-grain sublimation, i.e., $T_{\rm{ex}}$ higher for N-COMs than O-COMs.}
\vspace{-0.3cm} 
\tablenotetext{f}{Method used to derive the excitation temperatures. Here we only make a broad distinction between rotational diagram analysis (RD) or spectral modeling (M).}
\vspace{-0.3cm} 
\tablenotetext{g}{Total number of entries.}
\vspace{-0.3cm} 
\tablenotetext{h}{Total number of entries showing a sign of carbon grain sublimation.}
\end{deluxetable*}
\end{longrotatetable}

% ..........................................................

% ======================================================================
% REFERENCES
% ======================================================================

\bibliography{References}{}
\bibliographystyle{aasjournal}

\end{document}